# The search and study of PMS stars with Hα emission

Elena H. Nikoghosyan, Ani V. Vardanyan, and Knarik G. Khachatryan

*Byurakan Astrophysical Observatory, Byurakan vil., Aragatsotn reg., Armenia; elena@bao.sci.am*

**Abstract.** One of the most prominent features of young stellar objects in the optical range is the presence of emission lines, in particular Hα at 6563Å. Therefore, Hα emission detection is the most common spectroscopic means for identifying young stars. We present the search's results of PMS stellar objects in several star forming regions carried out on 2.6 m telescope in Byurakan observatory. We have used the method of slit-less spectroscopy employing a grism in combination with a narrow-band Hα interference filter to detect the objects with H α emissions.

## 1. Introduction

One of the most prominent features of young stellar objects in the optical range is the presence of emission lines, in particular the tracer of ionized gas Hα emission at 6563Å.

Already at the beginning of the last century near some dark nebulae it was discovered the stars of relatively late spectral classes with bright emission lines of hydrogen and ionized calcium. But only in 1945 American astrophysics Alfred Joy (Joy 1945) have proved, that these stars are not exotic anomaly, but the new class of objects, named T Tauri type stars and they should look for the presence of a bright hydrogen emission line, which is one of their main observational features.

Progress in understanding the evolutionary status of these stars was outlined very fast. Already at the end of 40th years famous Armenian astrophysicist Victor Ambartsumyan, on the basis of stellar dynamics in associations, suggested their youth (Ambartsumian 1949). A few years later it was also published papers, in which the basic characteristics of T Tauri stars were given, namely: 1) these stars occur in groups, forming T-associations; 2) in their spectra observed continual and lines emission; 3) they reveal strong variability (Ambartsumian 1954).

Later, the first catalogs were appeared. Most notable among them is the work of Guillermo Haro where the list of 250 Hα emission stars observed in an area 3.5 square degree located the brightest part of the Orion Nebula is presented (Haro 1953).

A large contribution to the study of this new class of stars made famous American astrophysicist George Herbig, who found many new stars of this type, and analyzed all the available observational evidence. In 1962 he published the first union catalog which contains the basic parameters of T Tauri stars, namely: 1) Sp from G to M; 2) Hα and Ca H, K emission, 3) Fe I 4063 and Fe I 4132 emission (Herbig 1962). But even before, G. Herbig was defined another class of young stellar objects, which are essentially intermediate analogous of TTau stars. Later they were named after the author Ae/Be Herbig stars. The author proposed for them the following characteristics: 1) Sp earlier than F0; 2) the emission lines of the Balmer series; 3) connection with nebula. As in the previous case, one of the main characteristics is Hα emission (Herbig 1960). Subsequently, G. Herbig and his coauthors were found about 1000 stellar objects with Hα emission in following star forming regions ONC, NGC 2264, NGC 1579, NGC 7000, IC 5070, IC 348, IC 2068, Messier 8 and 20 and etc (Reipurth 2008).

The studies of stars with Hα emission have always been in the focus of Byurakan observatory. Among the most significant works the following should be noted the studies in Cygnus, Orion and Taurus star forming regions of E. Parsamyan and her collaborations (Kazaryan & Parsamyan 1971; Parsamian 1981; Parsamian & Chavira 1982; Parsamian & Hojaev 1985). They not only revealed more than 500 emission stars in Orion Nebula, but also investigated the variability of Hα emission. The data, obtained on 40" Schmidt telescope allowed to N. Melikian and A. Karapetyan revealed and detailed study more than 200 emission stars in NGC 7000, IC 5068, IC 5070, NGC 6910, Cepheus and Cygnus star forming regions (Melikian & Shevchenko 1990; Melikyan 1994; Melikian & Karapetian 1996, 2010; Melikian et al. 2014). In the low-dispersion spectroscopic plates of the First Byurakan Survey K. Gigoyan and A. Mickaelyan with collaborations detected about 1000 Late-Type Stars with Hα emission (Rossi et al. 2011; Gigoyan & Mickaelian 2012).

Presently, in the list of the most studied for that matter star forming regions includes ONC, Taurus-Auriga, Perseus, Cygnus, Chepheus, Monoceros, Gum Nebula, Lagoon Nebula, North America and Pelican Nebulae, Ophiuchi cloud, NGC 2264 and L1228 areas (Reipurth 2008). The search of PMS stars with Hα emission it has gone already beyond our Galaxy. The authors of Reid & Parker (2012) paper by combining photometric and spectroscopic observations were able to identify about 600 emission stars in Large Magellanic Clouds.

Currently, the sources of information on the Hα activity of the young stellar population are large-scale surveys. The most prominent photometric surveys for search the Hα emission stars are both the IPHAS (Drew et al. 2005) and its southern counterpart VPHAS (Groot et al. 2009) surveys. These surveys cover about a 360 x 10 square degree view of the entire Galactic Plane at roughly one arcsec spatial resolution. In total, the present catalogue contains 4853 point sources that exhibit strong photometric evidence for Hα emission. Undoubtedly, a significant contribution in the study of Hα activity of PMS stars has already been invested and will be more invested the data obtained by Gaia survey. Already by using the sample of 22035 spectra the 3765 stars with intrinsic Hα emission have been found and classified the profile of emission line (Traven et al. 2015).

**2. The search of PMS stars with Hα emission by slit-less method**
For detection the stellar objects with Hα emission are used both photometric and spectral selections. In the latter case, the most widely used an objective prism or slit-less methods.

For our study we used the slit-less method, which has some advantages over the photometry and objective prism as will be discussed below. It is the combining of the grism working in the 5500-7500 Å range with a dispersion of 2.1 Å /pixel or 1.2 Å /pixel with a narrow-band Hα interference filter ( $\lambda c$ = 6560 Å, $\lambda$ = 85 Å). In the prime focus of the 2.6-m telescope of BAO with 1800 sec exposure it is possible with high certainty to find emission for the stars with R ≤ 21.0. The limit of the measuring of EW(Hα) is about 2 Å. For objects with R < 17.0 the measurement errors for EW(Hα) not more than 30%, for weaker objects error increases up to 40% of the EW(Hα). On the Fig.1 the example of the image was obtained by slit-less observation for IC 348 young stellar cluster is presented.

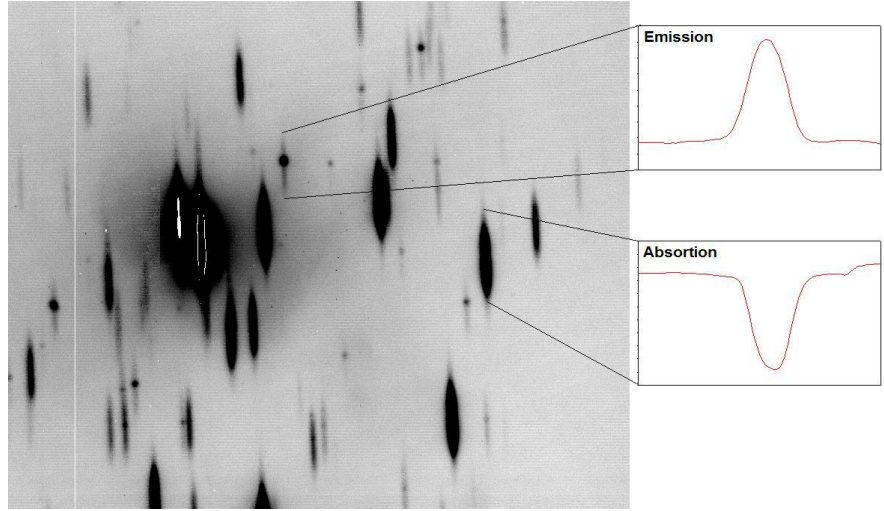

**Figure 1**. The sample of Hα slit-less method, applied in IC 348 stellar cluster

During the period from 2002 up to 2015 our group has done a number of works, in which were presented the results of the search and study of the stars with Hα emissions in several star forming regions, namely L 1340 (Magakian et al. 2003), NGC 7129 (Magakian et al. 2004), LkHα 326 (Movsessian et al. 2008), GM 1-64 and GM 2-4 (Nikogossian et al. 2009), GM 2-41 (Nikoghosyan et al. 2012), Cep OB3 (Nikoghosyan 2013), IC 348 (Nikoghosyan et al. 2015). Totally, we revealed about 250 emission stars and 200 of them are new discovered (see Table 1).

**Table 1. The number of the revealed stars with Hα emission**

| Region | N of revealed objects | N of new discovered |
|---|---|---|
| L 1340 | 14 | 11 |
| NGC 7129 | 22 | 16 |
| LkHα 326 | 6 | 5 |
| GM1-64 & GM 2-4 | 12 | 12 |
| Ceo OB2 | 43 | 30 |
| Cep OB3 | 150 | 123 |
| IC 348 | 196 | 5 |

As it was mentioned at the beginning, the slit-less method has some advantages. For example, by using objective prism in L 1340 and NGC 7129 on the images with same photometric limit, only 3 (Kun et al. 1994) and 4 (Hartigan & Lada 1985) emission stars were obtained correspondingly. In GM 2-41 region by photometric selection were revealed only 14 objects (Vink et al. 2008).

## 3. The main results
### 3.1. The numerical ratio between CTTau and WTTau objects

With respect to EW(Hα) the low-mass PMS stellar objects are conditionally divided into two main groups. The first one is CTTau objects, in which Hα is mainly produced as a result of accretion activity. The second group includes the WTTau objects which Hα emission mainly resulting from chromospheric solar like flares. Supposedly, these objects have a later evolution stage or III evolution Class, while CTTau objects belong to earlier II evolution Class. The standard limit value of EW(Hα) for these two groups of PMS stellar objects with different evolution classis is 10 Å. Besides, for the separation of the PMS objects according of their Hα activity also have been used the classifications proposed by White & Basri (2003), according to which, the limit value of EW(Hα) for WTTau and CTTau objects depends on their spectral classes. The critical limit changes from 3 Å for Sp earlier than K5 up to 20 Å - for Sp latter than M3.

In Table 2 the numerical ration between CTTau and WTTau objects obtained for our four regions, as well as age and distance module plus interstellar absorption of their parent cluster are presented. For comparison, we proposed the same data of other star forming areas in which the search of Hα emission stars carried out with about the same limit on both photometry and spectroscopy.

**Table 2. The CTTau/WTTau ratio, age and distance module**

| Region | CTTau : WTTau | Age (years) | Dist. mod. + $A_R$ |
|---|---|---|---|
| NGC 7129 | 67% : 33% | $3.0 \cdot 10^6$ | 11.8 |
| Cyg OB2 | 83% : 17% | $1.0 \cdot 10^6$ | 12.8 |
| Cep OB3 | 34% : 66% | $7.4 \cdot 10^5$ | 11.0 |
| IC 348 | 34% : 66% | $2.0 \cdot 10^6$ | 11.4 |
| For comparison | | | |
| L 988 (Herbig&Dahm, 2006) | 60% : 40% | $6.0 \cdot 10^5$ | 10.9 |
| LkHα 101 (Herbig et al, 2004) | 60% : 40% | $5.0 \cdot 10^5$ | 17.2 |
| NGC 2264 (Dahm, 2008) | 50% : 50% | $1.1 \cdot 10^6$ | 11.1 |
| IC 5146 (Herbig&Dahm, 2002) | 31% : 69% | $1.0 \cdot 10^6$ | 12.8 |

We can see, that there is no noticeable dependence between relative number of C and W TTau objects and distance module. This fact indicates a lack of samples selectivity. Therefore, the lack of correlation between the percentage of objects with deferent evolution classes and clusters' age reflects the real situation that somewhat unexpectedly. Moreover, the data analysis of two richness clusters Cep OB3 and IC 348 shows, that there is no noticeable difference between the special distribution of CTTau and WTTau stars according the direction of the star forming wave.

### 3.2. The correlation between Hα emission and other signs of activity

The existence of stellar envelope and disk, stellar wind and bipolar outflow determines the observational evidences of PMS objects, which cover the wavelength range from X-ray to radio, and show a wide range of phenomena, which distinguish PMS objects from other stellar population. The quantitative ratio of these features changes during the evolution from a very young protostars to the evolution class III.

The data in the Table 3 show the quantitative estimation of CTTau and WTTau objects in relation to the other signs of activity of PMS stars for Cep OB3 and IC 348 clusters. One of these sings is infrared excess characterized by the slope of the spectral energy distribution ($\alpha$) in the mid-infrared range, which is strongly depends on the evolutionary class of PMS stars. We can see, that there is the correlation between the value of infrared excess and H$\alpha$ activity. The majority of the objects with strong H$\alpha$ emission, as well as most of the stars of Cep BO3 for which EW(H$\alpha$) were not measured because of the weakness of the continuum, have a gently sloping spectral energy distribution in infrared range, scilicet high infrared excess, which can be explained by the presence of an optically thick disk and envelope components. And, on the contrary, most of the stars with faint emission and absorption in Cep OB3 region belong to later disk-less evolutionary classes.

Table 3. H$\alpha$ emission and other signs of activity

| Objects | N | X-ray | $\alpha$>-1.8 | -2.56<$\alpha$<-1.8 | $\alpha$<-2.56 | Age (years) |
|---|---|---|---|---|---|---|
| Cep OB 3 | | | | | | |
| CTTau | 39 | 25 (64%) | 39 (100%) | - | - | $4.6 \cdot 10^5$ |
| WTTau | 75 | 69 (92%) | 24 (32%) | 9 (12%) | 42 (56%) | $7.1 \cdot 10^5$ |
| Faint cont. | 34 | 12 (38%) | 23 (68%) | 6 (18%) | 5 (14%) | $4.8 \cdot 10^5$ |
| Absorption | 57 | 54 (95%) | 5 (9%) | 7 (12%) | 45 (79%) | $1.6 \cdot 10^6$ |
| IC 348 | | | | | | |
| CTTau | 77 | 25 (32%) | 65 (84%) | 8 (10%) | - | $2 \cdot 10^6$ |
| WTTau | 138 | 89 (64%) | 29 (21%) | 102 (74%) | - | $2 \cdot 10^6$ |
| no em. (M<1M$_\odot$) | | | | | | $2 \cdot 10^6$ |
| no em. (M>1M$_\odot$) | | | | | | $7 \cdot 10^6$ |

The percent amount of X-ray sources among the WTTau stars is considerably higher than among the CTTau stars. A similar pattern has been observed in other young stellar clusters (Feigelson et al. 2007). However, X-ray emission is also observed for a large number of stars with strong emission. One explanation for this relationship is the following: X-ray emission in older WTTau objects is mainly a product of chromospheric activity, while among the CTTau stars it is caused by accretion activity, which also assumes the presence of a massive disk. On the other hand, a massive disk can absorb a significant fraction of the X-rays. This, rather than an absence of X-ray activity, is probably the reason, why the X-ray emission in a large number of the CTTau stars lies below the detection threshold. Perhaps, the presence of circumstellar disk and its different location relative to the line of sight may also explain some discrepancy between values of IR excess and EW(H$\alpha$).

The age of stellar objects in these two areas has been defined by the different ways. For stellar population in Cep OB3 it was determined according the infrared excess by the model proposed Robitaille et al. (2006). The age of PMS stars in IC 348 was determined concerning the isochrones constructed on the model proposed in D'Antona & Mazzitelli (1994) and Baraffe et al. (1998). We can see that in the first case the age of samples is different; it is increasing with decreasing H$\alpha$ activity. In IC 348 cluster the age estimations for CTTau, WTTau and absorption objects with small mass are the same.

### 3.3. The percentage of objects with Hα emission

An example of IC 348 cluster we try to estimate the total percentage of objects with Hα emission in respect to their luminosity and check what effect does the selectivity of observations. On the Fig. 2 are presented the luminosity functions for the all objects and stars with emission in both visible and correction by the absorption R magnitudes.

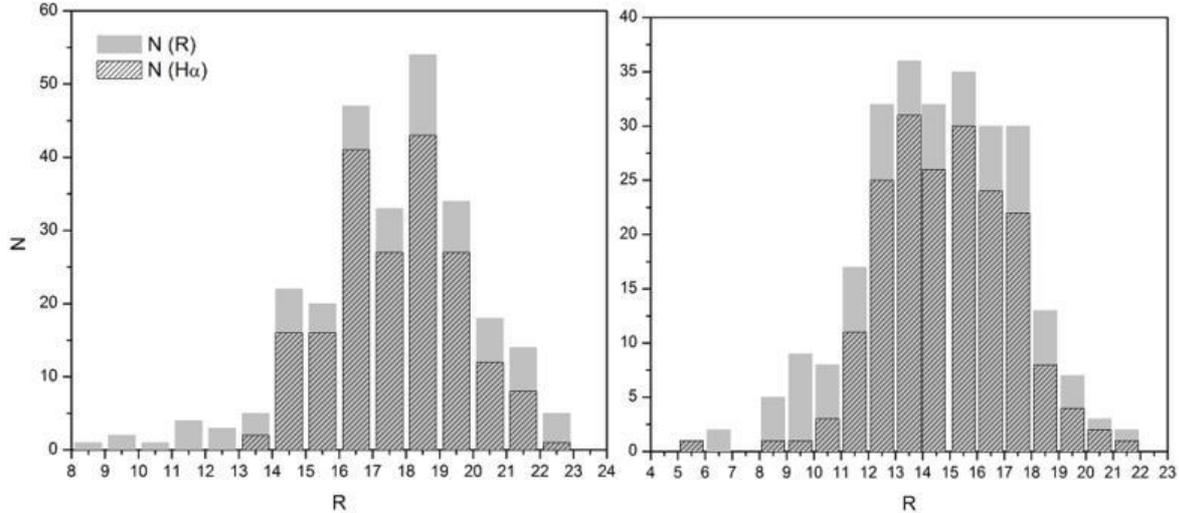

**Figure 2**. R luminosity function for visible R mag (left panel) and corrected for the absorption (right panel)

The histograms clearly show that the percentage of stars with Hα emission varies, depending on the magnitudes. On the left panel we can see, that the proportion of emission stars in the range $14.0 \leq R \leq 20.0$ reaches ~ 80% and remains practically unchanged. Among the bright stars with $R \leq 13.0$ no objects with Hα emission and among faint stars ($R > 20.0$) the percentage of emission stars is significant lower. The absorption correction does not significantly change the situation. Only by one magnitude shifted to the left the range with a relatively constant fraction of emission stars. In this case, it corresponds to the range of $13.0 \leq R \leq 19.0$. Apparently, 80% is the lower limit estimation of the relative number of stars with Hα activity. One possible reason for the lack of Hα emission in the remaining 20% is a variability between low emission and absorption, the close binaries, which can also be the cause of the measurement errors, which also cannot be excluded. Taking into account these factors, the percentage of emission stars certainly should raise and approach to 100%. The lower percentage of the emission stars among brighter likely reflection of reality. The decrease in the percentage of the emission objects among the fainter stars in the first, of course, can be explained by incompleteness of the observational data.

### 3.4. Conclusion

As a conclusion, we want to mention some not clear issues. If value of Hα activity, i.e. EW(Hα), represent the evolution stage of PMS objects, why:
- There is no noticeable dependence between the relative content of CTTau and WTTau stars and clusters' age.
- They are distributed evenly in cluster relative to the direction of the star formation

wave.
- There is no distinct difference in location of CTTau and WTTau stars according the evolution tracks.
- Can the position of the circumstellar disk with respect to the line of sight explains some discrepancy between the various manifestations of PMS activity (X-ray, Hα, IR excess)?